\documentclass{aastex631}

\usepackage[whole]{bxcjkjatype}

\begin{document}

\title{An ALMA-resolved view of 7000 au Protostellar Gas Ring around the Class~I source CrA-IRS~2 as a possible sign of magnetic flux advection}

\author[0000-0002-2062-1600]{Kazuki Tokuda}
\affiliation{Department of Earth and Planetary Sciences, Faculty of Science, Kyushu University, Nishi-ku, Fukuoka 819-0395, Japan}
\affiliation{National Astronomical Observatory of Japan, National Institutes of Natural Sciences, 2-21-1 Osawa, Mitaka, Tokyo 181-8588, Japan}

\author{Naofumi Fukaya}
\affiliation{Department of Physics, Nagoya University, Furo-cho, Chikusa-ku, Nagoya 464-8601, Japan}

\author[0000-0002-1411-5410]{Kengo Tachihara}
\affiliation{Department of Physics, Nagoya University, Furo-cho, Chikusa-ku, Nagoya 464-8601, Japan}

\author[0000-0002-7951-1641]{Mitsuki Omura}
\affiliation{Department of Earth and Planetary Sciences, Faculty of Science, Kyushu University, Nishi-ku, Fukuoka 819-0395, Japan}

\author[0000-0002-8217-7509]{Naoto Harada}
\affiliation{Department of Earth and Planetary Sciences, Faculty of Science, Kyushu University, Nishi-ku, Fukuoka 819-0395, Japan}

\author[0000-0003-4271-4901]{Shingo Nozaki}
\affiliation{Department of Earth and Planetary Sciences, Faculty of Science, Kyushu University, Nishi-ku, Fukuoka 819-0395, Japan}

\author[0000-0001-6580-6038]{Ayumu Shoshi}
\affiliation{Department of Earth and Planetary Sciences, Faculty of Science, Kyushu University, Nishi-ku, Fukuoka 819-0395, Japan}

\author[0000-0002-0963-0872]{Masahiro N. Machida}
\affiliation{Department of Earth and Planetary Sciences, Faculty of Science, Kyushu University, Nishi-ku, Fukuoka 819-0395, Japan}

\begin{abstract}

Transferring a significant fraction of the magnetic flux from a dense cloud core is essential in the star formation process. A ring-like structure produced by magnetic flux loss has been predicted theoretically, but no observational identification has been presented. We have performed ALMA observations of the Class~I protostar IRS~2 in the Corona Australis star-forming region and resolved a distinctive gas ring in the C$^{18}$O~($J$ = 2--1) line emission. The center of this gas ring is $\sim$5,000~au away from the protostar, with a diameter of $\sim$7,000~au. The radial velocity of the gas is $\lesssim1$\,km\,s$^{-1}$ blueshifted from that of the protostar, with a possible expanding feature judged from the velocity-field (moment~1) map and position-velocity diagram. These features are either observationally new or have been discovered but not discussed in depth because they are difficult to explain by well-studied protostellar phenomena such as molecular outflows and accretion streamers. A plausible interpretation is a magnetic wall created by the advection of magnetic flux which is theoretically expected in the Class~0/I phase during star formation as a removal mechanism of magnetic flux. Similar structures reported in the other young stellar sources could likely be candidates formed by the same mechanism, encouraging us to revisit the issue of magnetic flux transport in the early stages of star formation from an observational perspective.

\end{abstract}


\section{Introduction} \label{sec:intro}

The magnetic flux problem has long been a critical issue in star formation \citep[e.g.,][]{Nakano_1984}. If the magnetic fields observed in typical prestellar cores \citep{Troland_2008} were dragged into a young stellar object, the field strength would reach tens of millions of Gauss. This simple estimation is more than three orders of magnitude higher than the observed values, which are typically in the kilo-Gauss range \citep{Johns-Krull_2009}. Consequently, the majority of the magnetic flux of the prestellar core must be removed from the dense material infalling into the star, but when and how the removal occurs remains an unresolved issue.

A possible mechanism to remove magnetic flux is microscopic magnetic diffusion, such as ohmic dissipation and ambipolar diffusion, where charged particles frozen into the magnetic field drift through neutral particles, reducing the magnetic flux carried into the star and circumstellar disk \cite[e.g.,][]{Dapp_2012,Tomida_2015}. Although attempts have been made to observe this effect \citep{Yen_2018,Yen_2023}, it remains a challenging study. The ambipolar diffusion and ohmic dissipation decouple the neutral gas from the magnetic field in a steady-state manner over relatively long time scales. On the other hand, the magnetic interchange instability causes a more dynamic direct removal of magnetic flux from the circumstellar disk into its surroundings \citep{Parker_1979,Kaising_1992,Lubow_1995,Stehle_2001}.
During this time, outside the disk, a large amount of magnetic flux forms a low-gas density cavity that is surrounded by a ring-shaped dense gas cloud \citep{Zhao_2011,Machida_2014}, a structure sometimes referred to as a $``$magnetic wall$"$. These structures generally form on scales ranging from several tens to over 1,000\,au from the protostar \citep{Zhao_2011,Matsumoto_2017}. 
The magnetic field in the circumstellar disk can be transported to the edge of the disk due to ohmic dissipation and ambipolar diffusion, and the magnetic field strength (or magnetic flux) is enhanced at the edge. The magnetic interchange instability can occur when the ratio of magnetic strength to mass loading (or surface density), $B$/$\Sigma$, increases in the direction of gravity \citep{Lubow_1995}. This condition tends to be realized around the edge of the circumstellar disk due to magnetic diffusion within the disk \citep{Machida_2020}. After magnetic interchange instability occurs, the magnetic flux is rapidly leaked out from the disk when the magnetic pressure exceeds the ram pressure of the infalling envelope \citep{Krasnopolsky_2012}. 
While the occurrence of the magnetic interchange instability has been frequently demonstrated in numerical studies of protostar formation \cite[e.g.,][]{Joos_2012,Li_2013,Matsumoto_2017,Vaytet_2018,Machida_2020,Lee_2021}, there are also instances in simulations where the interchange instability does not arise due to ambipolar diffusion \citep{Masson_2016,Xu_2021a,Xu_2021b}. Theoretically, the manifestation of this instability in real protostellar systems remains ambiguous, and the lack of observational counterparts described in the next paragraph leads to it not being seriously considered until recently.

From an observational perspective, no structure produced by interchange instability has been reported in the literature. The advection of magnetic flux forms the ring-like or arc-like structure almost simultaneously with the protostar formation, and the size scales are very small, a few tens of au in the early stages, but grow into a larger structure on the order of the sound speed or free-fall time \citep[e.g.,][]{Zhao_2011,Matsumoto_2017}. Consequently, these scales could have been adequately detected with a spatial resolution available since the early scientific operation of ALMA toward nearby star-forming regions. Nevertheless, several reasons have prevented a successful identification. An embedded nature in the Class~0/I envelope makes discriminating the surrounding infalling and outflowing gas difficult. ALMA observations discovered complex arc-like envelopes around the Class~0 protostar in the Taurus dense core MC27/L1521F \citep{Tokuda_2014,Tokuda_2016,Tokuda_2018}, but the actual origin is still under debate. Only a few subsequent theoretical studies \citep{Matsumoto_2017,Machida_2020} discussed interchange instability as one of the possible physical mechanisms to reproduce the complex arc-like envelopes in MC27/L1521F. More recently, similar arc-like structures have been predominantly interpreted as accretion streams \citep[e.g.,][]{Pineda_2020}. Although the interchange-instability and streamer interpretations are not mutually exclusive, the simpler physical phenomenon of the latter reduces the opportunity to put the former on the table for discussion as an $``$observer$"$ bias.

In this latter, we report on ALMA observations toward the protostar IRS~2 in the Corona Australis (CrA) star-formation region whose distance is 149\,pc \citep{Galli_2020}. \cite{Forbrich_2007} classified it as a Class~I protostar with a spectral type K2. 
\cite{Sicilia-Aguilar_2013} performed the spectral energy distribution analysis of the source using multi-wavelength data. They determined the bolometric luminosity of $\sim$0.8\,$L_{\odot}$ and envelope temperature of $\sim$25\,K. Interestingly enough, their Herschel/PACS 100 and 160\,$\mu$m images show a bubble or shell-like structure with a diameter of $\sim$5,000\,au toward the southwest direction from the protostar (see Figure~4 in \citealt{Sicilia-Aguilar_2013}). Our ALMA data provide a detailed molecular gas view of the previously reported peculiar feature and an implication of the formation mechanism by the magnetic flux advection due to interchange instability, which may be an important factor in resolving the magnetic flux problem in the star formation process.

\section{Observations and Data Reduction} \label{sec:obs}

We performed ALMA Cycle~8 observations toward the CrA star-forming region (P.I., K. Tachihara; \#2021.1.00715.S) using the 12\,m array C43-1 configuration with the Band~6 receivers. The total mosaicing number of 137 provided the field of view of $\sim$180$\arcsec$ $\times$ 140$\arcsec$ (P.A. = 45\,deg.) at a central coordinate of ($\alpha_{\rm J2000.0}$, $\delta_{\rm J2000.0}$) = (19$^{\rm h}$01$^{\rm m}$39\fs4, $-$36\arcdeg58\arcmin16\farcs0). 
The main tracers used in this letter is C$^{18}$O~($J$ = 2--1). We supplementary used SO~($N,J$ = 5,6--4,5) and 1.3\,mm continuum emission to trace the gas and dust of the protostar vicinity. The C$^{18}$O and SO bandwidths were 59\,MHz with 480 channels at central sky frequencies of 219.562\,GHz and 219.951\,GHz, respectively. For the 1.3\,mm continuum, we used two spectral windows centered at 232.999\,GHz and 216.354\,GHz. The bandwidth and channel number were 1.875\,GHz and 1920, respectively, in each.

We used the Common Astronomy Software Application package \citep{CASAteam_2022}, v.6.4.4-31, in data reduction and imaging. We employed the \texttt{tclean} algorithm with the \texttt{multi-scale} deconvolver in the analysis. We applied the natural weighting. The clean mask regions were automatically determined by the \texttt{auto-multithresh} scheme \citep{Kepley_2020}. The resultant beam size and r.m.s. noise level of the C$^{18}$O and SO data are 2$\farcs$1 $\times$ 1$\farcs$3 (P.A. = $-$82\,deg.) and $\sim$0.2\,K at a velocity channel of 0.2\,km\,s$^{-1}$, respectively. For the 1.3\,mm continuum data, the beam size and sensitivity are 2$\farcs$0 $\times$ 1$\farcs$3 (P.A. = $-$78\,deg.) and $\sim$0.3\,mJy\,beam$^{-1}$, respectively. 

\section{Results} \label{sec:res}

\begin{figure}[htbp]
    \centering
    \includegraphics[width=1.05\columnwidth]{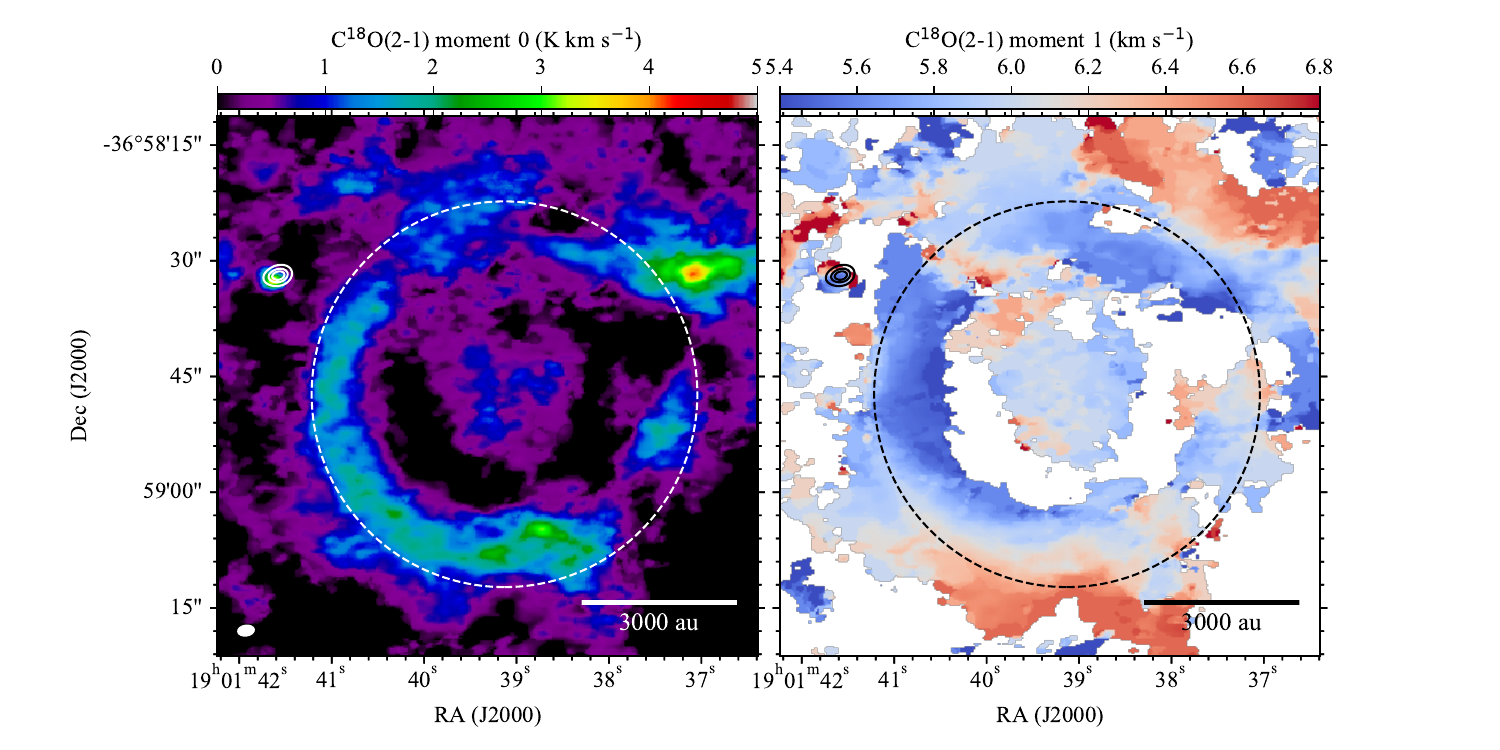}
    \caption{(Left panel) The color-scale image shows the velocity-integrated intensity (moment~0) map of C$^{18}$O~(2--1) with a velocity range of 5--7\,km\,s$^{-1}$. The white contours shows the 1.3\,mm continuum image with contour levels of 0.05, 0.15, and 0.25\,Jy\,beam$^{-1}$. The synthesized beam size, 2$\farcs$1 $\times$ 1$\farcs$3 is given by the white ellipse at the lower-left corner. The white dashed circle highlights the ring structure with a radius of 25$\arcsec$ ($\sim$3,700\,au) at a central coordinate of ($\alpha_{\rm J2000.0}$, $\delta_{\rm J2000.0}$) = (19$^{\rm h}$01$^{\rm m}$39\fs13, $-$36\arcdeg58\arcmin47\farcs3). (Right panel) The color-scale image shows the velocity-field (moment~1) map toward IRS~2. The contours and circle are the same as in the right panel.}
    \label{fig:mom0mom1}
\end{figure}

\begin{figure}[htb!]
    \centering
    \includegraphics[width=1.0\columnwidth]{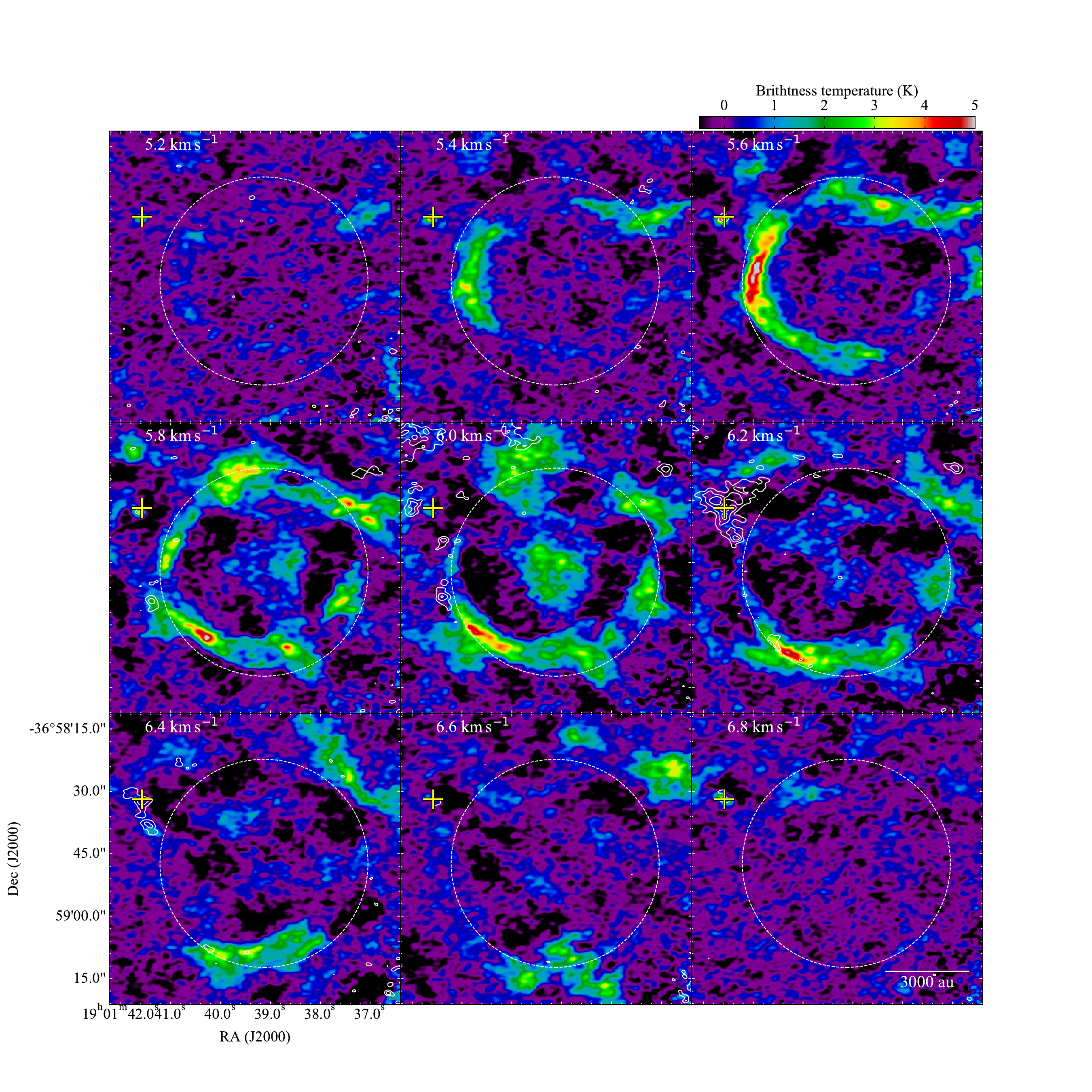}
    \caption{C$^{18}$O~(2--1) velocity-channel maps toward the ring structure around IRS~2 are shown in coloarscale images. The white lines are the same as those in Figure~\ref{fig:mom0mom1}. White contours show the SO~($N,J$ = 5,6--4,5) emission with contour levels of 0.9, 1.8, 2.7, and 3.6\,K. The yellow crosses demote the peak position of the 1.3\,mm continuum image, ($\alpha_{\rm J2000.0}$, $\delta_{\rm J2000.0}$) = (19$^{\rm h}$01$^{\rm m}$41\fs58, $-$36\arcdeg58\arcmin31\farcs9).}
    \label{fig:chanmap}
\end{figure}

Figure~\ref{fig:mom0mom1} shows the spatial and velocity distributions of 1.3\,mm continuum and C$^{18}$O~(2--1) emission toward IRS~2. The position of the continuum source corresponds to that of the previously identified protostar, tracing thermal dust emission arising from the circumstellar disk. A large ring-like structure is outstanding in the southwest direction of the protostar. In particular, because a continuous distribution of equal intensity of $\sim$1.5\,K\,km\,s$^{-1}$ follows the ring's southeast edge, we determined the ring structure guideline based on the boundary by eye as shown in the white dotted circle with a radius of 25$\arcsec$ ($\sim$3,700\,au). The projected separation is $\sim$35$\arcsec$ ($\sim$5,000\,au) between the protostar and the ring's center. Although the ring-like structure in the moment~0 and 1 maps is somewhat discontinuous in the northwest direction, the channel maps (Figure~\ref{fig:chanmap}) show that C$^{18}$O emission is visible along at least one of the velocity channels, according to the guidelines along the dotted circle. We regard it as a quasi-complete ring with a diameter of $\sim$7,000\,au around the protostellar source. 

\begin{figure}[htb!]
    \centering
    \includegraphics[width=0.8\columnwidth]{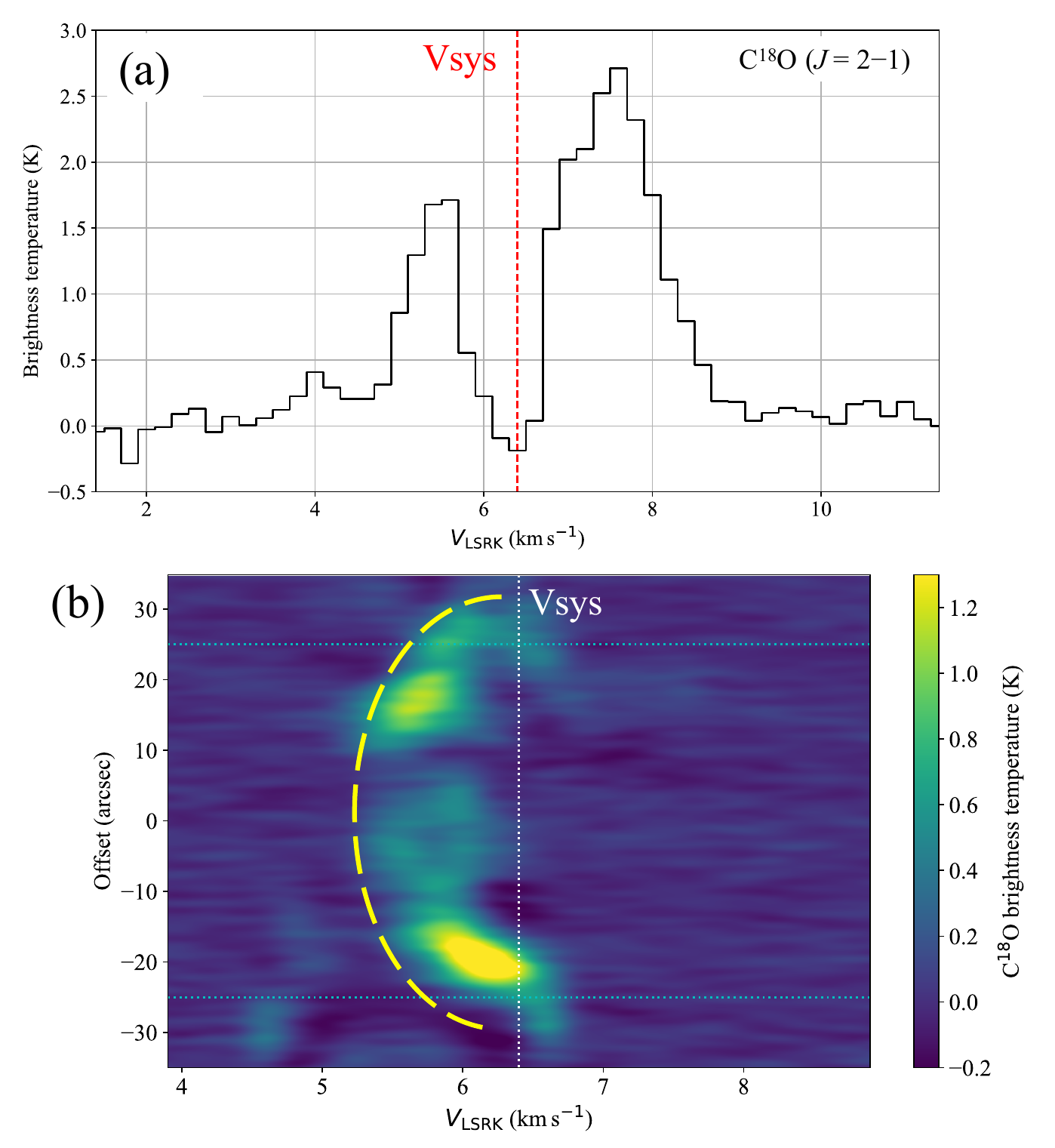}
    \caption{(a) An average C$^{18}$O spectrum toward IRS~2 over the region where the 1.3\,mm continuum mission is larger than 0.05\,Jy\,beam$^{-1}$. The red dashed line represents the systemic velocity, 6.4\,km\,s$^{-1}$ of the protostar judging from the dip velocity. (b) A C$^{18}$O position-velocity (PV) diagram toward over the ring structure. We extracted the spectral cube along the rectangle with a length of 70$\arcsec$ and a width of 55$\arcsec$ at the same central coordinate of the ring center as shown in Figure~\ref{fig:mom0mom1}. The y-axis shows the declination offset from the center. The cyan dotted lines along the horizontal axis denote the ring radius as shown in Figure~\ref{fig:mom0mom1}. The white dotted line represents the systemic velocity defined in panel (a).}
    \label{fig:spectPV}
\end{figure}

This ring was recognized as a continuous component with the protostar in the Herschel/PACS observations \citep{Sicilia-Aguilar_2013}. In addition to the C$^{18}$O data, we examined the SO emission in our ALMA dataset and found a distinct emission surrounding IRS~2 within a velocity range of 6.2--6.4\,km\,s$^{-1}$ (Figure~\ref{fig:chanmap}), closely matching the systemic velocity. Although C$^{18}$O emission around the systemic velocity is not visible (see Figure~\ref{fig:spectPV}a), possibly due to the optical thickness, combining it with the more optically thin SO distribution suggests that the ring structure and the protostar may constitute the same continuum component.

Based on the C$^{18}$O data, we obtained the average H$_2$ column density of $\sim$ 6 $\times$10$^{21}$\,cm$^{-2}$ along the ring assuming the local thermodynamical equilibrium (LTE) condition with a constant gas kinematic temperature of 25\,K \citep{Sicilia-Aguilar_2013} and applying a C$^{18}$O relative abundance, [H$_2$]/[C$^{18}$O], of 5.9 $\times$10$^{6}$, which is a typical value in the solar neighborhood \citep{Frerking_1982}. 
Suppose the ring thickness of $\sim$1,000\,au (100\,au), the average H$_2$ volume density is $\sim$4 $\times$10$^5$\,cm$^{-3}$ (4 $\times$10$^{6}$\,cm$^{-3}$). The total mass of the ring is roughly estimated to be $\sim$0.05\,$M_{\odot}$ with the integration of the column density map.

To investigate the dynamics of the ring, we examine its velocity structure. In Figure~\ref{fig:mom0mom1}b, there is a discernible blueshifted trend in the ring's inner edge and a subsequent redshift towards the outer. The channel maps, presented in Figure~\ref{fig:chanmap}, further elucidate this pattern. Gas components are more blueshifted than 5.7\,km\,s$^{-1}$ in the upper three panels located inside the ring. The C$^{18}$O emission is almost along the white dotted lines in the middle three panels (5.9--6.3\,km\,s$^{-1}$). 
At the bottom panels with the velocity of 6.5--6.7\,km\,s$^{-1}$, the primary emissions are extended toward outside the circle. 
Notably, around 6.1\,km\,s$^{-1}$ near the center, there is a prominent blob with an intensity of $\sim$3\,K, spanning more than 1,000\,au. Given its connection to the structure stretching northward, this component might be less a part of the ring and more likely an overlapping different cloud component along our line of sight. 

A noteworthy characteristic of this ring structure is overall blueshifted compared to the protostellar centroid velocity. To determine the systemic velocity of the protostar, we extracted the C$^{18}$O spectrum at the 1.3\,mm continuum disk position. Figure~\ref{fig:spectPV}a exhibits a dual-horned shape, indicative of a rotating motion \citep[e.g.,][]{Murillo_2013,Tokuda_2017}. The detailed velocity structure in channel maps or velocity diagrams of the disk itself is not visualized in this letter, but the direction of rotation is northwest-southeast, probably tracing the Keplerian motion. The dip is likely proximate to the protostar's systemic velocity \cite[e.g.,][]{Tokuda_2017}, estimated at 6.4\,km\,s$^{-1}$. As shown in Figure~\ref{fig:chanmap}, components redshifted more than 6.9\,km\,s$^{-1}$ are scarcely observed in the ring structure.

Figure~\ref{fig:spectPV}b shows a Position-Velocity (PV) diagram across the ring structure. The yellow curve highlighted in the diagram likely shows an expansion-like behavior that resembles expanding molecular or atomic gas components around H$\;${\sc ii} regions and supernova remnants \citep[e.g.,][]{Zhu_2008,Dawson_2008,Sano_2018,Sano_2021b}. Although the velocity distribution around driving sources depends on the initial gas configuration, in a scenario where the gas sweeps through a uniform medium, it would exhibit velocity structures symmetrical with respect to the central velocity. However, the IRS~2 ring structure only displays blueshifted components from the protostellar systemic velocity. This suggests that terms like expanding $``$bubble$"$ or $``$shell$"$ might not be the most fitting descriptions. Rather, referring to it as a $``$ring$"$ with an asymmetric velocity is more appropriate in the context.

To summarize the observed structures, we identified the high-density ($\sim$10$^{5}$\,cm$^{-3}$) gas ring with an expanding, blueshifted velocity gradient, approximately 7,000 au in diameter, positioned roughly 5,000 au from the protostar's center. Such characteristics appear to be largely unprecedented (or at least not clearly discussed) in recent observations of young protostellar sources.

\section{Discussion} \label{sec:dis}

We first discuss what exactly is the ring structure around the Class~I protostar, IRS~2, characterized in the present study. 
The CrA molecular cloud is a complex cluster-forming region with rich hierarchical molecular structures as a whole, and thus it cannot be completely ruled out that the ring structure is just a coincidence of unrelated line-of-site components along with the protostellar system as part of the natal cloud dynamics. Our forthcoming paper will discuss this possibility further (K., Tachihara et al. in prep.).  
The presence of the velocity gradient across the ring edge (Figures~\ref{fig:mom0mom1}b and \ref{fig:spectPV}b), which can be interpreted as an expanding motion, is reminiscent of the existence of an energy-driving source at the center, but no corresponding object has been reported so far (see also the next paragraph). The proximity of the observed velocity to that of the protostar, coupled with the distance of only a few thousand au, strongly suggests that the ring structure is likely related to IRS~2 directly. 
We proceed with our discussion, considering the ring as a structure or phenomenon associated with protostar formation hereafter.

Recent ALMA observations of protostellar systems have frequently reported what appear to be crescent or arc structures that are not pure rings but represent parts of them with different origins in each. \cite{Fern_2020} and \cite{Harada_2023} have identified ring or crescent-shaped molecular outflows that a pole-on configuration can explain. \cite{Sai_2023} detected an arc-like structure around the central protostar in the Ced110~IRS~4 System. The author's primary interpretation is a part of the outflowing gas. In the CrA IRS~2 system, the observed relative velocity of the C$^{18}$O ring with respect to the protostar is too small for the outflow interpretation, and we could not find the high-velocity component more than $\sim$5\,km\,s$^{-1}$. Even if there were a hidden protostar at the center of the ring driving an outflow at the currently observed lower limit velocity of $\lesssim$1\,km\,s$^{-1}$ (see Figure~\ref{fig:spectPV}b), its outflow force would be an order of $\sim$10$^{-6}$\,$M_{\odot}$\,km\,s$^{-1}$\,yr$^{-1}$ based on the ring mass of $\sim$0.05\,$M_{\odot}$ and the radius of $\sim$3,500\,au (see Sect.~\ref{sec:res}). Such an outflow-driving source should be brighter than the bolometric luminosity of $\sim$0.1--1\,$L_{\odot}$ \citep{Wu_2004}, which should be detectable with the currently available infrared survey, such as Spitzer.

The existence of arc-like structures, interpreted as accretion streamers, is garnering increasing attention \citep[e.g.,][]{Pineda_2020}. In the IRS~2 system, the SO component connected to the protostar is observed only around the systemic velocity (see Figure~\ref{fig:chanmap}). This characteristic differs from other potential streamers characterized by their velocity structure, which exhibits an accelerating velocity toward the protostar vicinities \cite[e.g.,][]{Harada_2023,Kido_2023}. In summary, the rings found in this IRS~2 system do not resemble structures or phenomena well known from previous observations of protostellar envelopes and are either newly discovered features or have not been discussed in depth.

We focus on the magnetic wall growth scenario induced by interchange instability (see Sect.~\ref{sec:intro}), reproduced in magnetohydrodynamics calculations as the mechanism generating the ring-like structure in the IRS~2 system. As illustrated in Figure~3 of \cite{Stehle_2001}, the interchange instability develops with a substantial azimuthal wave number, leading to the formation of multiple rings or holes in specific (off-center) directions. 
Although simulations tend to satisfy the conditions for repeatable interchange instability at the outer edge due to magnetic field dissipation within the disk, these conditions may not always be satisfied. The model in \cite{Machida_2020} appears to form two rings (see their Figure~5) from a single magnetic flux advection event because they imposed an initially axisymmetric density distribution. If the surrounding environment exhibits a non-uniform gas density around the disk, it is expected that the advection direction of the magnetic flux can be determined by the anisotropic ram pressure \citep{Krasnopolsky_2012,Matsumoto_2017}, resulting in the formation of a single ring \citep{Zhao_2011}. In this case, magnetic flux is leaked from the region where the ram pressure is weakest outside the disk. From an observational perspective, the primary filament in the CrA region is located northeast of IRS~2 \citep{Sicilia-Aguilar_2013}. This location is opposite the ring, suggesting the presence of an inhomogeneous gas distribution. A small-scale ($\sim$1,000\,au) gas density inhomogeneities prior to star formation are indeed observed in the Taurus region, where isolated dense cores are clearly identified \citep{Tokuda_2020Tau}. Therefore, the possibility of a singular ring formation by interchange instability cannot be excluded.

The magnetic-wall ring formed by interchange instability is created almost simultaneously with protostar formation with a size scale of a few tens au \cite[e.g.,][]{Joos_2012,Machida_2014,Matsumoto_2017,Machida_2020}. 
The advection of the magnetic flux produces the gas cavity that expands until the magnetic pressure balances with the ram pressure of the infalling gas, where a magnetic wall (high-density gas region) forms. The expansion time scale of the ring is roughly determined by the sound speed or on the order of the free-fall time \citep{Zhao_2011,Matsumoto_2017}. For example, in the IRS~2 system, if we assume that the ring expansion started immediately after the protostar formation at a constant speed of $\sim$0.3\,km\,s$^{-1}$, which is the sound speed at a temperature of 25\,K \citep{Sicilia-Aguilar_2013} and consistent with the observed relative velocity with respect to the protostar (Figures~\ref{fig:mom0mom1}b and \ref{fig:spectPV}b), it would take on the order of $\sim$10$^5$ years to reach the most distant ring edge from the protostar, $\sim$9,000\,au. Considering that the statistically derived timescale for Class~0 is $\sim$10$^{5}$\,years \cite[e.g.,][]{Evans_2009,Maury_2011}, the evolutionary stage of IRS~2 is consistent with the ring expansion estimate above if we assume that the ring and protostar formed simultaneously. While theoretical calculations suggest that rings could be deformed due to interactions within the infalling and rotating envelope (e.g., Figure~6 of \citealt{Zhao_2011}), the interchange instability-driven structure indicated in Figiure~2 of \cite{Zhao_2011} appears fairly circular as in the case of a non-rotating collapse case. Although prestellar cores are generally thought to be rotating in their initial conditions, it is conceivable that when the magnetic field is strong enough to induce interchange instability, the magnetic braking can suppress rotation, potentially rendering scenarios similar to cases without rotation. As mentioned in the previous section, the primary filament is in the opposite direction of the ring. Because the protostar is located at the edge of the dense parental cloud, it is natural that flux leaks in a low-density gas region and creates a cavity structure in the same direction without a significant perturbation from the infalling gas. Based on the current evidence, we conclude that the ALMA-resolved gas distribution of the IRS~2 ring is likely a product of the advection of the magnetic flux due to the interchange instability. Note that it is necessary to conduct additional theoretical calculations in the future to replicate the phenomena discussed here and verify which structures develop under various initial conditions that are more specialized to the situation of CrA IRS~2.

The structure discovered here may have some common characteristics in previously reported protostellar envelopes. For instance, arc-like structures with a size scale of $\sim$2,000\,au were discovered toward MC27/L1521F (see Sect.~\ref{sec:intro}). The arc's column density and velocity dispersion (see Table~2 in \citealt{Tokuda_2018}) are also consistent with those in the IRS~2 ring. The author's interpretation is that the complex arcs are formed by turbulent gas motion or dynamical interaction among multiple protostars and/or gas condensations (see also \citealt{Matsumoto_2015}). The lack of observational confidence to discuss the interchange instability was due to the ring's incomplete distribution. Such imperfect ring structures are also found in other Class~0 objects, such as VLA~1623 \citep{Mercimek_2023} and IRAS~16293-2422 \citep{Murillo_2022}. The observational examples cited here bear some resemblance to the features of the IRS~2 gas ring, providing value for revisiting their origin, although individual detailed studies are needed.

If the interchange-instability-driven ring expands as the function of the time, the density likely decreases, making it more likely detectable with low-density tracers such as $^{12}$CO. Possibly, the $^{12}$CO filamentary gas observed toward B59-BHB2007 \citep{Alves_2020} might correspond to the diffuse remnant of the ring. Identifying the interchange-instability origin ring likely becomes more challenging as the structures mix with the ambient mediums in the later evolutionary stages \citep{Zhao_2011}. Nevertheless, we propose that the potential to deepen our understanding of the protostellar evolution process may be hidden in the interaction between ring structures and the surrounding envelope, creating complex structures. 

The presence or absence of these rings may be vital in exploring the star formation magnetic flux problem, necessitating a dual approach from both observational and theoretical aspects. If we wish to prove the validity of interchange instability as the ring formation mechanism in CrA IRS~2, a stronger magnetic field must be observed within the hole. If the C$^{18}$O-traced ring is observed in lines causing the strong Zeeman splitting, such as CN and CCS, it may be worth challenging polarization observations to measure the magnetic field strength along the line-of-sight. Unfortunately, the Stokes~$I$, i.e., total line strength, is also expected to be weak in the hole, preventing a plausible detection of polarized emission with the current capabilities of ALMA. If the goal is to detect the magnetic field structure resulting from the interchange instability, targeting regions with newly formed rings without much expansion and with high column densities that show a thermal dust continuum may be a strategic approach. A higher polarization fraction in the dust continuum emission is indeed observed in arc-like protostellar envelopes \citep[e.g.,][]{Cox_2018,Takahashi_2018}, suggesting the presence of magnetic walls. However, the specific polarization pattern could vary depending on the evolutionary stage and viewing angle. It would be prudent to perform synthetic observations based on current numerical studies to understand how interchange instability-driven rings will be observed.
Future works to reveal the magnetic field could further clarify arc or ring's characteristics, enabling distinctions from different physical origin features such as accretion streamers.

\section{Summary}\label{sec:summary}

We presented ALMA observations of the Class~I protostar IRS~2 in the CrA star-forming region at a spatial resolution of $\sim$200\,au. We identified a large molecular gas ring, $\sim$7,000\,au in diameter, showing signs of expansion. The characteristics of this ring, specifically its size and location with respect to the protostellar source, are consistent with the theory of magnetic wall expansion caused by magnetic flux leakage due to interchange instability. Our observational results could represent the first preliminary evidence of magnetic flux leakage during the early stages of star formation, providing crucial insights into the star formation process, particularly regarding the magnetic flux problem. The ring's morphology, especially its highly circular shape, is a point of ongoing consideration. Future theoretical studies focusing on the dynamics of interchange-instability-driven structures under various conditions are crucial. Additionally, follow-up magnetic field observations and case studies involving other protostellar sources are essential for gaining a deeper understanding.


\begin{acknowledgments}
We would like to thank the anonymous referee for useful comments that improved the manuscript. This paper makes use of the following ALMA data: ADS/JAO. ALMA\#2021.1.00715.S. ALMA is a partnership of ESO (representing its member states), the NSF (USA), and NINS (Japan), together with the NRC (Canada), MOST, and ASIAA (Taiwan), and KASI (Republic of Korea), in cooperation with the Republic of Chile. The Joint ALMA Observatory is operated by the ESO, AUI/NRAO, and NAOJ. This work was supported by a NAOJ ALMA Scientific Research grant Nos. 2022-22B, Grants-in-Aid for Scientific Research (KAKENHI) of Japan Society for the Promotion of Science (JSPS; grant No. JP21K13962). 
\end{acknowledgments}

\appendix

%



\software{astropy \citep{Astropy18}, CASA \citep{CASAteam_2022}}



\end{document}